\documentstyle[prl,tighten,aps,epsfig,graphics,preprint,amstex]{revtex}
\newcommand{\ket}[1]{| \, #1 \rangle}
\newcommand{\bra}[1]{\langle \, #1 | }
\newcommand{\id}{\openone}

\begin{document}
\draft
\title{Quantum Remote Control: Teleportation of Unitary Operations}
\author{S.F. Huelga$^*$, J.A. Vaccaro and A. Chefles}
\address{Department of Physical Sciences, University of Hertfordshire, Hatfield AL10 9AB,
England}
\author{M.B. Plenio}
\address{Optics Section, The Blackett Laboratory,
Imperial College, London SW7 2BW, England}
\date{\today}
\maketitle
\begin{abstract}
We consider the implementation of an arbitrary unitary operation {\em U} upon a distant
quantum system. This teleportation of {\em U} can be viewed as a {\em quantum remote
control}. We investigate protocols which achieve this using local operations, classical
communication and shared entanglement (LOCCSE).  Lower bounds on the necessary
entanglement and classical communication are determined using causality and the linearity
of quantum mechanics. We examine in particular detail the resources required if the remote
control is to be implemented as a classical black box.  Under these circumstances, we
prove that the required resources are, necessarily, those needed for implementation by
bidirectional state teleportation.
\end{abstract}
Much of the current fascination with quantum information
processing derives from the properties of entanglement
\cite{martin}. On one hand, entanglement can give rise to nonlocal
correlations which defy explanation in terms of local, realistic
theories\cite{Bell}, but on the other, it can also be used as a
resource. While it is impossible, for example, to determine the
state of a quantum system, entanglement makes it possible to
transmit an unknown state. This process is known as quantum state
teleportation \cite{tele}. Quantum state teleportation can be
linked directly to various interrelated principles of quantum
information processing, such as the impossibility of superluminal
communication, the non-increasing of entanglement under local
operations and classical communication \cite{martin} and the
no-cloning theorem \cite{woot}. However, the information contained
in the state of a quantum system is only one kind of information
which is important in quantum mechanics.  Another is the
information which describes quantum operations.  In this paper, we
examine the issue of teleporting, not a quantum state, but rather
a quantum operation. In particular, we examine the teleportation
of an unknown unitary operation on a qubit. This procedure would
function in a manner similar to that of a remote control
apparatus, and so we shall also refer to it as {\em quantum remote
control}. We will first pose the problem in a completely general
theoretical framework and focus later on an experimentally
feasible scenario where entanglement resources are limited.\\ The
most general scenario for the teleportation of an arbitrary
unitary operation is depicted in Figure 1. One party, Alice,
possesses a physical system, {\em C}, which we shall refer to as
the {\em control}. The control contains information describing a
unitary operation {\em U} upon the state of a qubit, and is itself
a quantum system. The control state corresponding to the unitary
operation {\em U} will be denoted by $\ket{U}_{C}$. Her colleague
Bob has a qubit ${\beta}$ prepared in the state
$\ket{\psi}_{\beta}$. The aim is to devise a physical procedure
which effects the transformation
$\ket{\psi}_{\beta}{\mapsto}U\ket{\psi}_{\beta}$, for every
initial state $\ket{\psi}_{\beta}$ and every unitary operation
$U$.  The most general such procedure can be represented by a
completely positive, linear, trace preserving map on the set of
density operators for the combined $C{\beta}$ system. Any such map
has a unitary representation $\cal T$ involving ancillary systems.
We shall denote the state of the ancilla at Alice's and Bob's
laboratories by $\ket{\chi}_{AB}$. Then the teleportation
operation has the general form
\begin{equation}
{\cal T} \big[\ket{\chi}_{AB}{\otimes}\ket{U}_{C}{\otimes}\ket{\psi}_{\beta}\big]=
\ket{\Phi(U,{\chi})}_{ABC}{\otimes}\left(U\ket{\psi}_{\beta}\right). \label{tony}
\end{equation}
In the following we investigate some of the properties of ${\cal
T}$. In particular, we derive lower bounds on the amount of
non-local resources that are needed to implement ${\cal T}$ using
only local operations and classical communication. The unitary
teleportation operator ${\cal T}$ is independent of both $U$ and
$\ket{\psi}_{\beta}$. The final state of the ancilla+control,
$\ket{\Phi(U,{\chi})}_{ABC}$, must be independent of
$\ket{\psi}_{\beta}$.  To see why, let us suppose that it isn't,
in which case there will be at least one {\em U}, and two states,
$\ket{\psi}_{\beta}$ and $\ket{\psi^{'}}_{\beta}$, for which
$\ket{{\Phi}(U,{\chi},{\psi})}_{ABC}{\neq}\ket{{\Phi}(U,{\chi},{\psi}^{'})}_{ABC}$.
We imagine that {\em U} is successfully teleported for the states
$\ket{\psi}_{\beta}$ and $\ket{\psi^{'}}_{\beta}$.  Suppose now
that Bob's qubit is prepared in a superposition of these states,
$(c_{1}\ket{\psi}+c_{2}|{\psi}^{'}{\rangle})_{\beta}$. The
linearity of ${\cal T}$ implies that the final total state will be
\begin{equation}
({\id}_{ABC}{\otimes}U_{\beta})\big{[}c_{1}\ket{\Phi(U,{\chi},{\psi})}_{ABC}{\otimes}
\ket{\psi}_{\beta}+c_{2}\ket{\Phi(U,{\chi},{\psi}^{'})}_{ABC}{\otimes}\ket{{\psi}^{'}}_{\beta}\big{]}.
\end{equation}
The requirement that Bob's qubit undergoes a unitary evolution
implies that it must not be entangled with the remaining systems.
However, one can see that it is entangled with ABC whenever
$c_{1}c_{2}{\neq}0$.  Thus, the final state of ABC must be
independent of $\ket{\psi}_{\beta}$.\\ The set of all unitary
operations {\em U} is infinite. This implies that if the dimension
of the control system is to be finite, then the control states
$\ket{U}_{C}$ must, in general, be non-orthogonal. However,
Nielsen and Chuang showed, in a slightly different context, that
this cannot be the case \cite{uni}. The problem investigated by
these authors was whether or not one could devise a universal
programmable quantum gate array, which could be used to store and
execute any program upon a quantum register. They showed that no
such finite array can be constructed. Their method of proof can
readily be transferred to this context, making use of the
correspondences between programmable gate array/control, and
register/Bob's qubit. Following their reasoning, we note that Eq.
(1) and the unitarity of ${\cal T}$ imply that, for any two
different unitary transformations $U$ and $U'$,
\begin{equation}
\frac{{}_{C}\bra{U'}U{\rangle}_{C}}{{}_{ABC}\bra{{\Phi}(U',{\chi})}{\Phi}(U,{\chi}){\rangle}_{ABC}}={}_{\beta}\bra{\psi}U^{'\dagger}U\ket{\psi}_{\beta}.
\end{equation}
The left hand side is independent of $\ket{\psi}_{\beta}$, and this equality is true for
all $\ket{\psi}_{\beta}$.  It follows that $U^{'\dagger}U={\gamma}{\id}$, for some
constant ${\gamma}$, leading to the conclusion that $U$ and $U'$ are identical up to a
multiplicative constant. This conclusion, however, is valid only when the denominator on
the left hand side is non-zero.  If it is zero, then ${}_{C}\bra{U'}U{\rangle}_{C}=0$, by
the unitarity of ${\cal T}$. Control states corresponding to different unitary
transformations are orthogonal, so that no finite-dimensional control system can be used
to teleport an arbitrary unitary operation. For the remainder of this paper, when we speak
of an arbitrary unitary operation, we will mean one which belongs to some arbitrarily
large, but finite, set.  We will also assume that this set contains the identity
$\sigma^0={\id}$ and the 3 Pauli operators ${\sigma}^{i}$. Note that the orthogonality of
the control states opens the possibility that different operations can, at least in
principle, be distinguished by Alice. This will only be possible though if Alice knows the
basis $\{|U\rangle_C\}$ in which the information is encoded.

The teleportation of $U$ is a collective operation on spatially
separated systems, which we wish to carry out using shared
entanglement and classical communication. In the derivation of
lower bounds on the amount of non-local resources that are
required to implement the teleportation of $U$ locally, two
guiding principles will be very useful \cite{martin}:

\noindent{\bf (i)} {\em The amount of classical information able
to be communicated by an operation in a given direction across
some partition between subsystems cannot exceed the amount of
information that must be sent in this direction across the same
partition to complete the operation.}

\noindent {\bf (ii)} {\em The amount of bipartite entanglement that an
operation can establish across some partition between subsystems cannot exceed
the amount of prior entanglement across the partition that must be consumed in
order to complete the operation.}

We now use principle (i) to establish the fact that at least two classical bits must be
sent from Alice to Bob to complete the teleportation of an arbitrary {\em U}.  Suppose
that, rather than being prepared in a pure state, Bob's qubit is initially maximally
entangled with some other qubit, ${\beta}^{'}$, which is also in Bob's laboratory.  Let us
denote the four Bell states for a pair of qubits by $\ket{B^{\mu}}$, where
${\mu}=0,{\dots},3$. Using the technique of super-dense coding\cite{dense}, any of the
four Bell states can be transformed into any other by application of one of the Pauli
operators ${\sigma}^{i}$ on one of the qubits. We take this qubit to be ${\beta}$, and
notice that the $\ket{B^{\mu}}$ can be ordered in such a way that
$({\sigma}^{\mu}_{\beta}{\otimes}{\id}_{{\beta}'})\ket{B^{0}}_{{\beta}{\beta}'}=\ket{B^{\mu}}_{{\beta}{\beta}'}$.
Alice can easily transmit two bits of information to Bob if he prepares the
${\beta}{\beta}'$ system in the state $\ket{B^{0}}_{{\beta}{\beta}'}$.  She chooses the
control system to be in one of the states $\ket{{\sigma}^{\mu}}_{C}$.  Following the
action of ${\cal T}$, Bob will be in possession of the corresponding Bell state
$\ket{B^{\mu}}_{{\beta}{\beta}'}$. If he subsequently performs a Bell measurement on
${\beta}{\beta}'$, then he will be able to determine the value of ${\mu}$, and hence the
control state which Alice prepared, revealing 2 bits of classical information.\\
We now show that, by teleporting an arbitrary {\em U} according to the general
prescription in Eq. (\ref{tony}), Alice and Bob can establish 2 ebits of shared
entanglement. Imagine that, in addition to the systems we have already introduced, Alice
has a further 4-dimensional ancilla, which we shall label {\em R}. Let the states
$\ket{{\mu}}_{R}$ be a particular orthonormal basis for {\em R}. Suppose now that Alice
initially prepares {\em R} and the control {\em C} in the maximally entangled state
$(1/2)\sum_{\mu}\ket{{\mu}}_{R}{\otimes}\ket{{\sigma}^{\mu}}_{C}$. Bob once more prepares
the Bell state $\ket{B^{0}}_{{\beta}{\beta}'}$.  The teleportation operation ${\cal T}$ is
then carried out according to Eq. (\ref{tony}). It is more convenient here, however, to
work with a form of this equation that represents, explicitly, any local measurements made
by Alice and Bob and any classical communication between them. In this case ${\cal T}$ in
Eq. (\ref{tony}) is replaced by a pair of classically-correlated local CP maps, one in
each laboratory.  Classical information is revealed by measurements, and we let the index
{\em i} denote each measurement outcome.  The final state corresponding to the {\em i}th
outcome is
\begin{equation}
\ket{{\psi}_{F}}_i=\frac{1}{2}\sum_{\mu}\ket{{\mu}}_{R}{\otimes}
  \ket{\Phi_i({\sigma}^{\mu},{\chi})}_{ABC}{\otimes}\ket{B^{\mu}}_{{\beta}{\beta}'}
\end{equation}
We now calculate the entanglement shared by Alice and Bob. Alice is in possession of the
compound system {\em RAC}, while Bob has the system $B{\beta}{\beta}'$.  For each outcome,
these subsystems have respective density operators ${\rho}^{i}_{RAC}$ and
${\rho}^{i}_{B{\beta}{\beta}'}$. Since $\ket{{\psi}_{F}}_i$ is a pure state, it follows
that the entanglement shared by Alice and Bob is simply the (base 2) von Neumann entropy
of either of these density operators. Fortunately, we can calculate this explicitly. To do
so, we notice that the states $\ket{\Phi_i({\sigma}^{\mu},{\chi})}_{ABC}$ will generally
contain entanglement between {\em B} and {\em AC}.
Let us write ${\rho}^{i\mu}_{B}={\mathrm
Tr}_{AC}(\ket{\Phi_i({\sigma}^{\mu},{\chi})}\bra{\Phi_i({\sigma}^{\mu},{\chi})})$. We find
that\cite{footnote}
\begin{equation}
  {\rho}_{B{\beta}{\beta}'}=\frac{1}{4}\sum_{\mu}(\ket{B^{\mu}}\bra{B^{\mu}})_{{\beta}{\beta}'}
   {\otimes}{\rho}^{i\mu}_{B}.
\end{equation}
Making use of the orthogonality of the $\ket{B^{\mu}}$, we find that the total entropy of
entanglement shared by Alice and Bob is simply
\begin{equation}
E(\ket{{\psi}_{F}})=S({\rho}_{B{\beta}{\beta}'})=2+\frac{1}{4}\sum_{\mu}S({\rho}^{i\mu}_{B})\,
{\geq}\, 2.
\end{equation}
It follows from principle (ii) that at least 2 ebits of
entanglement need to be consumed to implement ${\cal T}$ locally,
i.e. to teleport an arbitrary unitary operation.

We can summarize the results obtained so far as follows.  The resources required to
perform quantum remote control can be classified into shared entanglement, classical
information transmission from Alice to Bob, and from Bob to Alice. We have established
absolute lower bounds on the first two of these resources. Alice and Bob have to share at
least two ebits and Alice needs to transmit to Bob, at least, two bits of classical
information.

These bounds can be attained by a procedure in which Bob teleports the state of his
particle to Alice who, after applying the unitary transformation, teleports it back to
him. We will call this the ``bidirectional state teleportation'' scheme. The scheme
requires sending 2 classical bits in each direction, and using 2 ebits of shared
entanglement. It would also be conceivable to adopt a different strategy -- teleporting
the state of the control system from Alice to Bob who would then implement the control
directly onto $\beta$. We call this the ``control-state teleportation'' scheme.

Control-state teleportation is a unidirectional communication scheme from Alice to Bob, so
the absolute lower bound for the communication exchange from Bob to Alice is zero.
Obviously, the overall resources will depend on the dimensionality of the control system
$C$. We cannot say anything about the optimality of this procedure; whether there exists
another unidirectional protocol which uses less resources is an open problem.

On the other hand, bidirectional state teleportation saturates the
lower bounds for the amount of shared ebits and classical bits
transmitted from Alice to Bob and additionally uses two bits of
classical communication from Bob to Alice. This scheme allows the
faithful implementation of $U$ independently of the dimension of
the control system.  To be more efficient overall, any other
scheme would need less resources than bidirectional state
teleportation. This establishes an upper bound in the overall
amount of resources required for the efficient remote
implementation of an arbitrary $U$ as 4 classical bits and 2
ebits.

We now consider an experimental scenario where the black box
implementing an arbitrary transformation $U$ is a macroscopic
object, involving a (very) large number of degrees of freedom. The
option of teleporting the control apparatus is then unfeasible,
given that it would consume an infinite amount of entanglement and
classical communication resources. However, the question remains
whether there exists a more economical protocol than bidirectional
state teleportation. We will prove in the following that this is
not possible and bidirectional state teleportation is an
unconditional optimal way to remotely implement an arbitrary $U$.

Discarding the possibility of control-state teleportation allows us to replace the
transformation given by Eq. (\ref{tony}) with
\begin{equation}
  G_2 \, U \, G_1 (\ket{\chi}_{\alpha AB} \otimes\ket{\psi}_{\beta})
   =  \ket{\Phi(U,\chi)}_{\alpha AB} \otimes U \ket{\psi}_{\beta}
  \label{doble},
\end{equation}
where certain fixed operations $G_1$ and $G_2$ are performed,
respectively, prior to and following the action of the arbitrary
$U$ on a qubit $\alpha$ on Alice's side. We assume that Alice and
Bob share initially some entanglement, represented by the state
$\ket{\chi}_{\alpha AB}$. As before, the purpose of the
transformation is to perform the operation $U$ on Bob's qubit
$\beta$. We continue to use a nonlocal unitary representation of
the transformation where $G_1$ and $G_2$ are unitary operators
acting on possibly all subsystems. A pictorial scheme of the
situation using a quantum circuit is given in Figure 2. The two
upper wires refer to Alice's subsystems and the two lower ones to
Bob's. Note that operations $G_i$ are represented by non-local
gates while the action of $U$ takes place locally on Alice's side.

We prove in the following that the only way that Eq. (\ref{doble}) can be implemented
(locally) is by teleporting the state $\ket{\psi}_\beta$ from Bob to Alice, and then
teleporting back the transformed state $U\ket{\psi}_\beta$ from Alice to Bob.

We begin by noting that linearity forces the transformed state of
systems $\alpha AB$ to be independent of the particular input
state $\ket{\psi}_{\beta}$.  In addition, linearity imposes the
condition that the state $\ket{\Phi(U,\chi)}_{\alpha AB}$ has to
be independent of {\em U} itself. To see this, consider the case
where the transformation $U$ is one of the four Pauli operators
$\sigma^\mu$ and assume that the global state of $\alpha AB$ after
completing the protocol may depend on the choice of $U$. According
to Eq. (\ref{doble}), the combined action of the operations $G_i$
has to be such that
\begin{equation}
G_2 \, \sigma^{\mu} \, G_1 \left( \ket{\chi}_{\alpha AB} \otimes
\ket{\psi}_{\beta} \right) = \ket{\Phi(\sigma^{\mu},\chi)}_{\alpha
AB}{\otimes} (\sigma^{\mu}\ket{\psi}_{\beta}). \label{inde}
\end{equation}
On the other hand, an arbitrary one-qubit unitary transformation
$U$ can always be decomposed in terms of the Pauli operators,
$U=\sum_{\mu=0}^4 \alpha_{\mu}\sigma^{\mu}$, and it must hold that
\begin{eqnarray}
  G_2 \, U \, G_1 \left(\ket{\chi}_{\alpha AB} \otimes\ket{\psi}_{\beta} \right)
      &=& \sum_{\mu} \alpha_{\mu} \ket{\Phi(\sigma^{\mu},\chi)}_{\alpha AB} \otimes
          (\sigma^{\mu}\ket{\psi}_{\beta}).
\end{eqnarray}
For the RHS to be a product state, as is required by Eq.
(\ref{doble}), we must have $\ket{\Phi(\sigma^{\mu},\chi)}_{\alpha
AB}=\ket{\Phi(\chi)}_{\alpha AB}$, independent of the operator
$\sigma^{\mu}$. This is true for any basis set of operators, and
so the final state of the ancillas $\alpha AB$ on the RHS of Eq.
(\ref{doble}) is independent of $U$.\\

We can now show that the operation $G_1$ necessarily has to be
non-trivial.  We do this by first assuming the contrary that
$G_1=\id$, and considering two input states, $\ket{\psi}_{\beta}$
and $\ket{\psi'}_{\beta}$ such that
$_{\beta}{\langle}{\psi}^{'}|{\psi}{\rangle}_{\beta}=0$, and two
unitary transformations ${\em U}$ and ${\em U'}$ which bring these
two states to the same state $\ket{\gamma}_{\beta}$.  Using Eq.
(\ref{doble}), this implies that
\begin{eqnarray}
  G_2 \left( U\ket{\chi}_{\alpha AB} \, \ket{\psi}_{\beta} \right)
      &=& \ket{\Phi(\chi)}_{\alpha AB}{\otimes}\ket{\gamma}_{\beta} \nonumber \\
  G_2 \left( U' \ket{\chi}_{\alpha AB} \, \ket{\psi'}_{\beta} \right)
      &=& \ket{\Phi(\chi)}_{\alpha AB}{\otimes} \ket{\gamma}_{\beta}\ .
  \label{chu}
\end{eqnarray}
No universal unitary action $G_2$ can be found to satisfy Eq.
(\ref{chu}), as this would require the mapping of orthogonal
states onto the same state. This shows that no universal operation
$G_2$ that satisfies Eq. (\ref{chu}) can exist and therefore, for
the $U$-teleportation to succeed, $G_1$ has to be non-trivial.

The final step in our proof is to rewrite Eq. (\ref{doble}) as
\begin{equation}
U \, G_1 (\ket{\chi}_{\alpha AB} \otimes \ket{\psi}_{\beta}) =
G^\dagger_2(\ket{\Phi(\chi)}_{\alpha AB} \otimes U \ket{\psi}_{\beta}). \label{doble2}
\end{equation}
Since $G_1$ and $G_2$ are universal gates, we may choose $U$ and $\ket{\psi}_\beta$
freely. For each $\ket{\psi}_\beta$ let the operator $U_\psi$ be such that
$U_\psi\ket{\psi}=\ket{0}$ where ${\sigma}_{z}\ket{0}=\ket{0}$.  If
$U={\sigma}_{z}U_{\psi}$, then
\begin{eqnarray*}
  \left(\sigma_z U_\psi\right) G_1 \left(\ket{\chi}_{\alpha AB}
              \otimes\ket{\psi}_\beta\right)
  &=& G^\dagger_2\left(\ket{\Phi(\chi)}_{\alpha AB}\otimes
              \sigma_z U_\psi \ket{\psi}_\beta \right)\\
  &=& G^\dagger_2\left(\ket{\Phi(\chi)}_{\alpha AB}\otimes\ket{0}_\beta\right)\ .
\end{eqnarray*}
The RHS is simply $(U_\psi) G_1\left(\ket{\chi}_{\alpha
AB}\otimes\ket{\psi}_\beta\right)$ and so, necessarily,  $(U_\psi)
G_1\left(\ket{\chi}_{\alpha AB}\otimes\ket{\psi}_\beta\right)$ is
the eigenstate $\ket{0}_{\alpha} \otimes\ket{\phi}_{AB \beta}$ of
$\left(\sigma_z\right)_{\alpha} \otimes\id_{AB \beta}$.
Equivalently,
\begin{eqnarray}
  G_1\left(\ket{\chi}_{\alpha AB}\otimes\ket{\psi}_\beta\right)
  &=& \left(U^\dagger_\psi\ket{0}_\alpha\right)\otimes\ket{\phi}_{AB \beta} \nonumber\\
  &=& \ket{\psi}_\alpha\otimes\ket{\phi}_{AB \beta} \label{tele}\ .
\end{eqnarray}
In other words, the operation $G_1$ necessarily transfers Bob's
state $\ket{\psi}$ to Alice's qubit $\alpha$.  Substituting Eq.
(\ref{tele}) into Eq. (\ref{doble}) then shows that $G_2$
necessarily transfers $U\ket{\psi}$ back to Bob's qubit ${\beta}$.
From these results and the fact that quantum state teleportation
is an optimal procedure for local state transfer, we conclude that
the optimal procedure for implementing locally a universal {\em
U}-teleportation scheme is by means of bidirectional state
teleportation.\\
In this paper we have investigated the potential use of LOCCSE for the remote control of a
quantum system. We have determined requirements that must be satisfied by any method that
implements this task by LOCCSE means. In particular, we have shown that, if Alice can
teleport an arbitrary unitary operation to a qubit in her colleague Bob's laboratory, then
she must communicate at least two bits of classical information to him, and they must
share at least 2 ebits of entanglement. If the unitary operation is remotely implemented
by a classical apparatus, then to effect the teleportation at least 2 classical bits must
also be transmitted from Bob to Alice. These resources can be used to perform the
teleportation of {\em U} using bidirectional state teleportation. Remarkably, no protocol
employing a smaller amount of resources is possible.\\ We believe that this work will
stimulate further research into ways in which LOCCSE can be used to control remotely the
properties of other quantum system, with potential applications ranging from remotely
synchronized time evolutions to distributed quantum computing.
\\

{\bf Acknowledgements:} The authors thank O. Steuernagel and S. M. Barnett for
discussions and D. Jonathan and S. Virmani for critically reading the
manuscript. This work has been supported by The Leverhulme Trust,
the EQUIP project of the European Union, the Engineering and Physical Sciences
Research Council (EPSRC) and DGICYT Project No. PB-98-0191 (Spain).\\ $(*)$
Permanent Address: Departamento de F\'{\i}sica. Universidad de Oviedo. Calvo
Sotelo s/n. 33007 Oviedo. Spain.

\newpage
\begin{figure}[tbp]
\epsfxsize6.5cm
 \centerline{\epsfbox{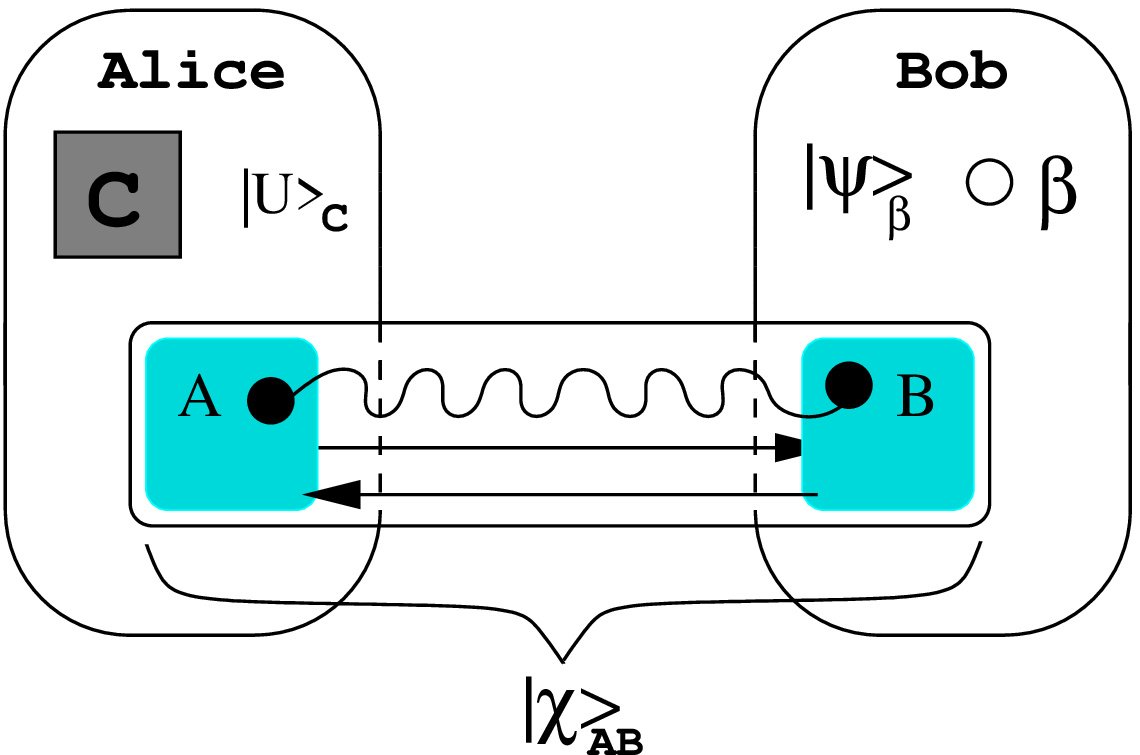}}
 \vspace*{0.2 cm}
 \end{figure}
Figure 1. Caption.\\ Initial setup involved in the teleportation
of an arbitrary unitary operation. The control system C in Alice's
laboratory is initially prepared in the state $\ket{U}_{C}$,
corresponding to the unitary operation {\em U}.  This operation is
to be remotely carried out on Bob's qubit ${\beta}$, which is
initially prepared in an arbitrary pure state
$\ket{\psi}_{\beta}$.  This will be achieved by local operations
in the individual laboratories, involving a collective ancilla
initially prepared in the state $\ket{\chi}_{AB}$, supplemented by
the exchange of classical communication, represented in the
diagram by the arrow lines.\\

\begin{figure}[hbt]
\epsfxsize8.5cm \centerline{\epsfbox{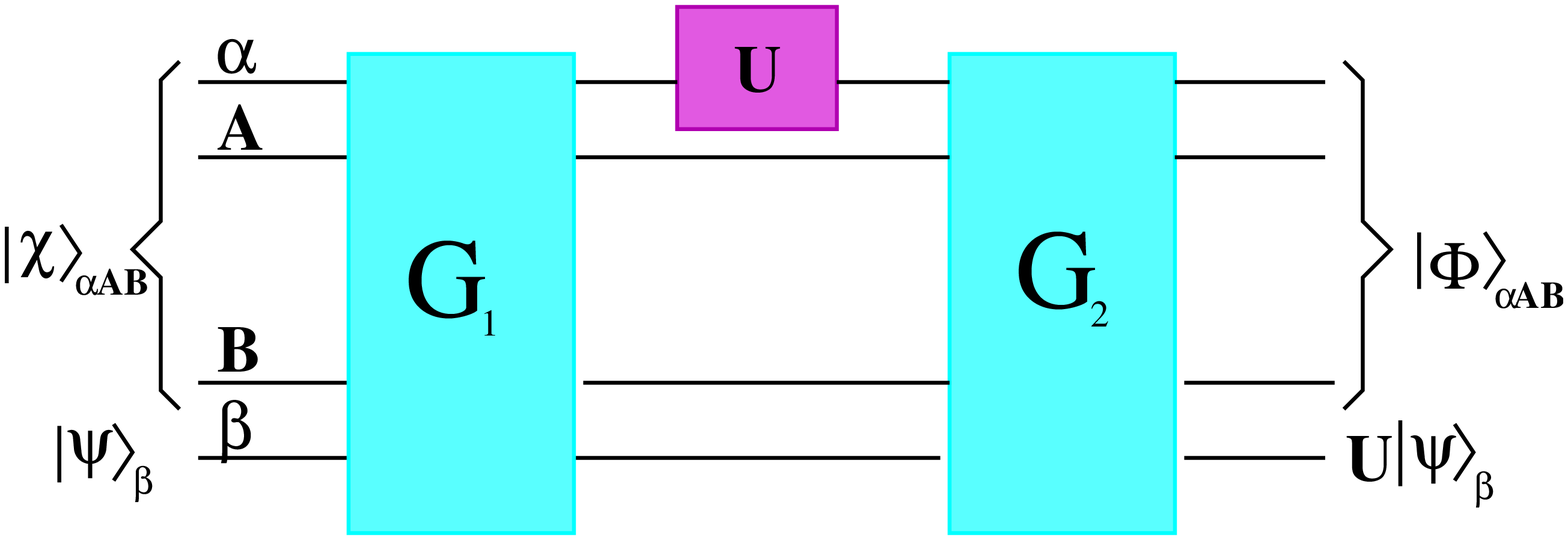}} \end{figure}
Figure 2. Caption.\\
 Quantum circuit representation of the process of
teleporting an arbitrary one-qubit transformation. The two upper
wires belong to Alice and the lower ones to Bob. Initially Alice
and Bob share some entanglement, represented by the joint state
$\ket{\chi}_{\alpha AB}$. Operations $G_1$ and $G_2$ are modeled
in terms of non-local unitary transformations.

\begin{thebibliography}{99}
\bibitem{footnote}
Although subsystems may move from one laboratory to the other for the purposes of
classical communication, their locations will be revealed by the measurement, making
${\rho}_{B{\beta}{\beta}'}^{i}$ well defined.
\end{thebibliography}
\end{document}